\documentclass[10pt,twocolumn]{article}
\usepackage{makeidx}
\usepackage{mathrsfs,amsmath}
\usepackage{amsfonts}
\usepackage{amssymb}
\usepackage{setspace}
\usepackage{cite}
\usepackage{ol2}
\usepackage{multicol}
\usepackage{epstopdf}
\usepackage{graphicx}
\usepackage{caption}

\begin{document}

\twocolumn[
\title{Breathing solitary-pulse pairs in a linearly coupled system}
\author{Brenda Dana, Boris A. Malomed and Alon Bahabad}
\address{Department of Physical Electronics, School of Electrical Engineering,
Fleischman Faculty of Engineering, Tel-Aviv University, Tel-Aviv
69978, Israel}

\begin{abstract}
It is shown that pairs of solitary pulses (SPs) in a
linearly-coupled system with opposite group-velocity dispersions
form robust breathing bound states. The
system can be realized by temporal-modulation coupling of
SPs with different carrier frequencies propagating in the same
medium, or by coupling of SPs in a dual-core waveguide. Broad SP pairs are
produced in a virtually exact form by means of the
variational approximation. Strong nonlinearity
tends to destroy the periodic evolution of the SP pairs.

\end{abstract}

 \maketitle
 ]


\textit{Introduction and the model}. Breathing solitary pulses (SPs) emerge
as fundamental modes in various optical media, including fiber lasers \cite%
{Orenstein}-\cite{Kutz}, systems based on the periodic dispersion management
(DM) \cite{malomed2006soliton}-\cite{Ganapathy} or nonlinearity management
\cite{Barcelona}, and oscillations of spatial solitons in external traps
\cite{Koby,Serkin}. In those settings, the breathing dynamics is usually
induced by the periodic structure of the system. In this work, we
demonstrate that robust breathing SP pairs emerge in uniform systems built
as linearly coupled guided modes with opposite group velocity dispersions
(GVDs). In fact, completely stable breathing SP pairs can be created without
nonlinearity. If the nonlinearity is too strong, it actually destroys the
pair.

We start with a system of two coupled linear Schr\"{o}dinger equations,
which is a particular case of mixed discrete-continuous systems \cite{LSE}:

\begin{equation}
\frac{\partial A_{j}}{\partial z}+v_{gj}^{-1}\frac{\partial A_{j}}{\partial t%
}-\frac{i\beta _{j}}{2}\frac{\partial ^{2}A_{j}}{\partial t^{2}}=i\kappa
_{0}A_{3-j},  \label{eqn:2}
\end{equation}%
$j=1,2$, with $v_{gj}^{-1}\equiv \frac{k(\omega _{j})}{b_{j}v_{g}(\omega
_{j})}$, $\beta _{j}\equiv \frac{k(\omega _{j})}{b_{j}}\frac{\partial
^{2}k(\omega )}{\partial \omega ^{2}}|_{\omega =\omega _{j}}$, where $%
v_{g}(\omega _{j})\equiv \left( {\frac{\partial k(\omega )}{\partial \omega }%
|_{\omega =\omega _{j}}}\right) ^{-1}\ $and $\frac{\partial ^{2}k(\omega )}{%
\partial \omega ^{2}}|_{\omega =\omega _{j}}$ are the group velocity and
GVD, respectively. Here $k(\omega )$ is the frequency-dependent wavenumber
for the system's normal modes, carried by eigenfrequencies $\omega _{j}$ and
respective propagation constants $b_{j}$ \cite{yariv2007photonics}.

These coupled-mode equations (CPEs) apply to different physical settings.
First, they may describe the co-propagation of two modes with the same
carrier frequency in a dual-core waveguide with different GVD coefficients
in the cores (due to different core materials or waveguide profiles), cf.
Refs. \cite{Kaup,Boardman}. In this case, the matching between $b_{1}$ and $%
b_{2}$ may be supported by an appropriate spatial modulation (e.g., in a
grating coupler \cite{yariv2007photonics}). A more promising possibility is
to embed a pair of waveguides into a photonic-crystal-fiber (PCF) matrix
\cite{PCF}. In the latter situation, the three conditions of the equality
between the phase and group velocities in the cores, and opposite GVD
coefficients, which are assumed below, can be secured using such parameters
as the diameter of the waveguides, the PCF pitch, and the carrier
wavelength.

Alternatively, the same equations govern the co-propagation of two modes in
the same waveguide, carried by different frequencies, with the matching
between $\omega _{1}$ and $\omega _{2}$ provided by a suitable temporal
modulation. Such temporal gratings are the subject of research in linear
\cite{PhysRevE.75.046607,1063-7869-48-8-R03} and nonlinear optics \cite{Alon}%
. Equations (\ref{eqn:2}) can be derived for these physical settings from
full CPEs which explicitly contain the spatial and temporal modulations as
mechanisms for the wavenumber and frequency matching \cite{brenda1}.

As said above, we focus on the symmetric system, with equal group velocities
and opposite GVD for the linearly coupled waves: $\beta _{2}=-\beta
_{1}\equiv \beta $,~$v_{g1}=v_{g2}\equiv v_{g}$. Equations (\ref{eqn:2}) are
then rescaled by defining $\xi \equiv \left( |\beta |/T_{0}^{2}\right) z$
and $\tau \equiv (t-z/v_{g})/T_{0}$, where $T_{0}$ is a characteristic
temporal width of the input pulse:

\begin{equation}
\frac{\partial A_{j}}{\partial \xi }-\frac{i}{2}(-1)^{j}\frac{\partial
^{2}A_{j}}{\partial \ \tau ^{2}}=iK_{0}A_{3-j}.  \label{eqn:4}
\end{equation}%
The corresponding SP period, normalized coupling coefficient, and coupling
length are $Z_{0}=\pi T_{0}^{2}/\left( 2|\beta |\right) $, $K_{0}=\kappa
_{0}T_{0}^{2}/|\beta |$ and $L_{c}\equiv \pi /\left( 2K_{0}\right) $,
respectively \cite{agrawal2010applications,yariv2007photonics}.
Equations (\ref{eqn:4}) can be derived from the Lagrangian density, with the
asterisk standing for complex conjugate:
\begin{align}
\mathcal{L}& =\left( i/2\right) (A_{1}^{\ast }A_{1\xi }-A_{1}A_{1\xi }^{\ast
}+A_{2}^{\ast }A_{2\xi }-A_{2}A_{2\xi }^{\ast })  \notag \\
& +(i/2)(|A_{1\tau }|^{2}-|A_{2\tau }|^{2})+K_{0}(A_{1}^{\ast
}A_{2}+A_{2}^{\ast }A_{1}).  \label{eqn:8}
\end{align}%
Note that the dispersion relation for plane-wave solution to Eq. ( \ref%
{eqn:4}), $A_{j}=A_{j}^{(0)}\exp \left( iQ\xi -i\Omega \tau \right) $ is
\begin{equation}
Q^{2}=K_{0}^{2}+\Omega ^{4}/4,  \label{gap}
\end{equation}%
hence solitary modes may exist in the respective spectral \textit{gap}, $%
Q^{2}<K_{0}^{2}$ \cite{Kaup}.

\textit{Analytical and numerical solution for the linear system}. The
commonly known fundamental solution of the single linear Schr\"{o}dinger
equation is a spreading Gaussian. In the case of DM with exactly vanishing
path-average GVD, $\bar{\beta}=0$, the solution is a breathing Gaussian
which does not suffer spreading. The Kerr nonlinearity stabilizes the
solution against spreading at $\bar{\beta}\neq 0$, giving rise to DM
solitons \cite{malomed2006soliton}-\cite{Ganapathy}.

In the present system, robust non-spreading breathing SPs are possible
without any DM, due to the action of the GVD terms with opposite signs,
linked by the linear coupling. An analytical solution can be constructed for
broad SPs, with $T_{0}^{2}\gg 1/K_{0}$
, using the variational approximation (VA) \cite{Chu:93,PhysRevA.41.6287}. A
relevant ansatz, describing rapid oscillations between the two modes, is:

\begin{equation}
\{A_{1}(\xi ,\tau ),A_{2}(\xi ,\tau )\}=A(\xi ,\tau )\left\{ i\sin (K_{0}\xi
),\cos (K_{0}\xi )\right\} ,  \label{eqn:6}
\end{equation}%
where $A(\xi ,\tau )$ is a slowly varying complex amplitude [i.e., these
solutions are looked for near edges of the above-mentioned spectral gap, see
Eqs. (\ref{gap})]. The ansatz is substituted into Eq.~(\ref{eqn:8}), keeping
only terms with derivatives of the slowly varying amplitude to yield an
effective Lagrangian density, $\mathcal{L}_{\mathrm{eff}}=(1/2)\left[
i(A^{\ast }A_{\xi }-AA_{\xi }^{\ast })-|A_{\tau }|^{2}\cos (2K_{0}\xi )%
\right] $. It gives rise to Euler-Lagrange equation with \emph{effective} DM
corresponding to $\bar{\beta}=0$, although no DM is present in the
underlying CPEs (\ref{eqn:2}):

\begin{equation}
iA_{\xi }+(1/2)A_{\tau \tau }\cos (2K_{0}\xi )=0.  \label{eqn:10}
\end{equation}%
%
%
%
%
Equation (\ref{eqn:10}) gives rise to exact Gaussian solutions for
non-spreading breathing SPs:

\begin{equation}
A(\xi ,\tau )=\frac{A_{0}T_{0}}{\sqrt{2T_{0}^{2}+K_{0}^{-1}\sin (2K_{0}\xi )}%
}\exp \left( -\frac{\tau ^{2}}{2T_{0}^{2}+iK_{0}^{-1}\sin (2K_{0}\xi )}%
\right) ,  \label{eqn:11}
\end{equation}


\noindent where the initial temporal width $T_{0}\gg 1/K_{0}$ and $A_{0}$
are arbitrary real constants.

\begin{figure}[th]
\begin{minipage}[b]{1\linewidth}
\includegraphics[width=\textwidth]{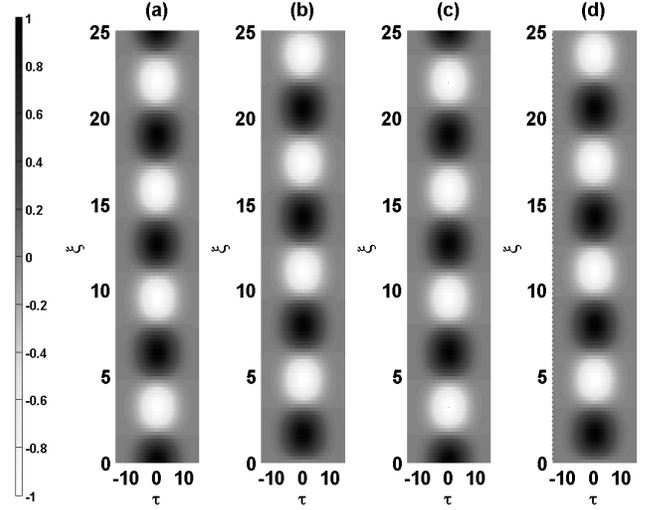}
\centering \caption{The field evolution in a broad SP
pair, with $K_{0}=1$, $T_{0}^{2}=20$. (a,b): The numerical solution of Eq. (%
\protect\ref{eqn:4}) for $\mathrm{Re}(A_{2}(\protect\xi ,\protect\tau ))$
and $\mathrm{Im}(A_{1}(\protect\xi ,\protect\tau ))$, respectively. (c,d):
Approximate analytical solution given by Eqs. (\protect\ref{eqn:6}) and (%
\protect\ref{eqn:11}) for the same components.}
\label{fig1:Fig1}
\end{minipage}
\end{figure}

\begin{figure}[th]
\begin{minipage}[b]{1\linewidth}
\includegraphics[width=\textwidth]{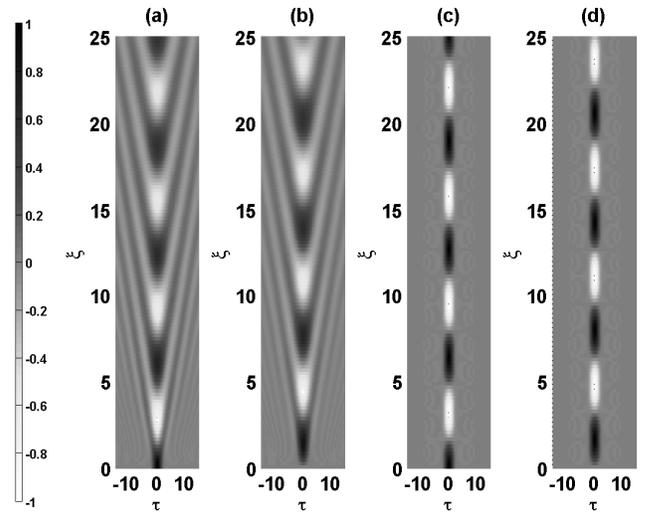} 
\end{minipage}
\caption{The same as in Fig. \protect\ref{fig1:Fig1}, but for a narrow SP
pair, with $T_{0}=1$.}
\label{fig2:Fig2}
\end{figure}

Thus, Eqs. (\ref{eqn:11}) and (\ref{eqn:6}) produce a breathing SP pair,
built as two components swinging at coupling frequency $K_{0}$, multiplied
by the common amplitude, $A(\xi ,\tau )$, oscillating at the double
frequency. In Figs. \ref{fig1:Fig1} and \ref{fig2:Fig2}, these approximate
analytical solutions are compared to results produced by the numerical
integration of Eqs. (\ref{eqn:4}) by means of the split-step
Fourier-transform method \cite{agrawal2010applications,Li20082811}.
It is seen that non-spreading SP solutions, oscillating almost precisely
with period $2\pi /K_{0}$, exist in all cases, the broad pulses being
perfectly approximated by the analytical solution, as expected, while for
narrow ones the approximation is inaccurate.

\noindent

\begin{figure}[th]
\centering
\begin{minipage}[b]{1\linewidth} \includegraphics[width=\textwidth]{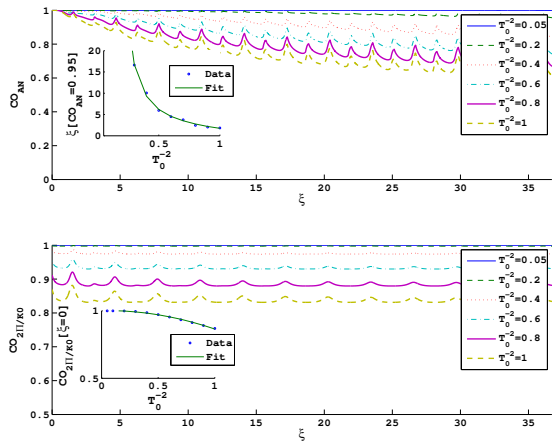}
\end{minipage}
\caption{Correlators computed in the linear system. (a) $\mathrm{CO}_{%
\mathrm{AN}}(\protect\xi )$ for different temporal widths $T_{0}$ of the SP
input. The inset: $\protect\xi (\mathrm{CO}_{\mathrm{AN}}=0.95)$ vs. $%
T_{0}^{-2}$, the continuous line showing a fit to an exponential function.
(b) $\mathrm{CO}_{2\protect\pi /K_{0}}(\protect\xi )$ for different $T_{0}$.
The inset: $\mathrm{CO}_{2\protect\pi /K_{0}}(\protect\xi =0)$ vs. $%
T_{0}^{-2}$, the continuous line showing a fit to a parabola.}
\label{fig3:Fig3}
\end{figure}

To analyze the solutions, we define the following correlator between two
pairs of functions $\{f_{1}(\xi ,\tau ),f_{2}(\xi ,\tau )\}$ and $%
\{g_{1}(\xi ,\tau ),g_{2}(\xi ,\tau )\}$:

\begin{equation}
\mathrm{CO}(\{f_{1},f_{2}\},\{g_{1},g_{2}\},\xi )=\frac{|\langle
f_{1}|g_{1}\rangle _{\tau }|+|\langle f_{2}|g_{2}\rangle _{\tau }|}{%
||f_{1}||_{\tau }||g_{1}||_{\tau }+||f_{2}||_{\tau }||g_{2}||_{\tau }},
\label{eqn:13}
\end{equation}%
with the inner product, $\langle f(\xi ,\tau )|g(\xi ,\tau )\rangle _{\tau
}\equiv \int_{-\infty }^{+\infty }{f^{\ast }(\xi ,\tau )g(\xi ,\tau )d\tau }$%
, and the corresponding norm, $||f||_{\tau }=\sqrt{\langle f|f\rangle _{\tau
}}$. The correlator takes values $0\leq \mathrm{CO}\leq 1$, with $\mathrm{CO}%
=1$ and $\mathrm{CO}=0$ corresponding, severally, to perfect correlation and
no correlation. Then, we can evaluate the proximity of the SP to the
periodic behavior as $\mathrm{CO}_{2\pi /K_{0}}=\mathrm{CO}(\{A_{1}(\xi
,\tau ),A_{2}(\xi ,\tau )\},\{A_{1}(\xi +2\pi /K_{0},\tau ),A_{2}(\xi +2\pi
/K_{0},\tau )\})$, and the consistency between approximate analytical and
numerical solutions: $\mathrm{CO}_{\mathrm{AN}}=\mathrm{CO}(\{A_{1\mathrm{N}%
},A_{2\mathrm{N}}\},\{A_{1\mathrm{A}},A_{2\mathrm{A}}\})$. In particular, as
shown in Fig. \ref{fig3:Fig3}(a), the latter correlator allows one to assess
\emph{how long} the approximate solution remains valid. Naturally, the
correlations decay with the increase of $\xi $ and decrease of $T_{0}$. The
inset shows the value of the propagation distance, $\xi $, at which $\mathrm{%
CO}_{\mathrm{AN}}$ drops to $0.95$, as a function $T_{0}^{-2}$, which
demonstrates an exponential decrease of the propagation range, in which the
analytical solution is valid, with the decrease of the SP's width.

Further, in Fig. \ref{fig3:Fig3}(b) the $\mathrm{CO}_{2\pi /K_{0}}$
correlator shows how well the SP pair maintains a periodic evolution
pattern. Increasing $T_{0}^{-2}$ causes a parabolic decrease in the
correlation, while the degree of the deviation from the perfectly periodic
behavior is itself periodic in $\xi $, oscillating at the double frequency, $%
2K_{0}$.

\textit{The nonlinear system}. It is straightforward \ to add Kerr terms to
the CPEs (\ref{eqn:4}) \cite%
{agrawal2010applications,yariv2007photonics,Li20082811}:%
\begin{equation}
\frac{\partial A_{j}}{\partial \xi }-\frac{i}{2}(-1)^{j}\frac{\partial
^{2}A_{j}}{\partial \ \tau ^{2}}=iK_{0}A_{3-j}+i\Gamma |A_{j}|^{2}A_{j},
\label{eqn:14}
\end{equation}

\noindent where $\Gamma $ is the scaled nonlinearity coefficient.
Accordingly, the Lagrangian density (\ref{eqn:8}) acquires an extra term, $%
\left( \Gamma /2\right) |\left( A_{1}|^{4}+|A_{2}|^{4}\right) $, and the VA
can be applied to the nonlinear system as well \cite%
{Chu:93,PhysRevA.41.6287,Li20082811}, again assuming $T_{0}\gg 1/K_{0}$.
Using the same ansatz (\ref{eqn:6}) as above leads to a nonlinear Schr\"{o}%
dinger (NLS) equation featuring a combination of DM with $\bar{\beta}=0$
[cf. Eq. (\ref{eqn:10})] and nonlinearity management: \noindent
\begin{equation}
iA_{\xi }+(1/2)A_{\tau \tau }\cos (2K_{0}\xi )+(\Gamma /4)\left[ \cos
(4K_{0}\xi )+3\right] |A|^{2}A=0.  \label{NLSE}
\end{equation}%
This equation can readily produce DM solitons solutions, by dint of methods
elaborated in the analysis of the DM and nonlinearity-managed systems \cite%
{malomed2006soliton}-\cite{Barcelona}.

\begin{figure}[th]
\centering
\begin{minipage}[b]{1\linewidth} \includegraphics[width=\textwidth]{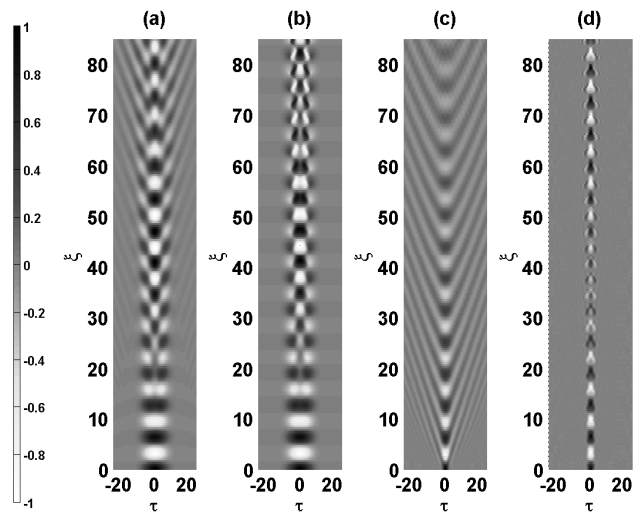}
\end{minipage}
\caption{The evolution of the $\mathrm{Re}\{A_{2}\}$ field in the nonlinear
system with $\Gamma =0.1$ for a broad SP with $T_{0}=20$: (a) a numerical
solution of the full system (\protect\ref{eqn:14}); (b) simulations of the
VA-produced single NLS equation (\protect\ref{NLSE}). (c,d): The same as in
(a,b), but for a narrow SP with $T_{0}=1$.}
\label{fig4:Fig4}
\end{figure}

Comparing numerical solutions produced by Eq. (\ref{NLSE}) with numerical
solutions for the SP pairs produced by simulations of the full system (\ref%
{eqn:14}) demonstrate that the strong nonlinearity tends to gradually
destroy the solitons. In Fig. \ref{fig4:Fig4} we display two representative
cases for $T_{0}=20$ and $T_{0}=1$ with $\Gamma =0.1$. Only the evolution of
$\mathrm{Re}\{A_{2}\}$ is shown, as it is sufficient to represent the
situation. The nonlinearity starts to affect the SP shape at propagation
distances exceeding the nonlinearity length, $\sim 1/\Gamma =10$. Similar to
the linear system (cf. Figs. \ref{fig1:Fig1} and \ref{fig2:Fig2}), the VA,
i.e., Eq. (\ref{NLSE}), is accurate for broad solitons, and inaccurate for
narrow ones. The gradual destruction of the SP by the nonlinearity is
naturally explained by the fact that, while the fundamental frequency of its
internal oscillations falls into the gap [see Eq. (\ref{gap})], higher-order
harmonics, generated by the cubic nonlinearity, couple to the continuous
spectrum, initiating decay of the SP.

\begin{figure}[th]
\centering
\begin{minipage}[b]{1\linewidth} \includegraphics[width=\textwidth]{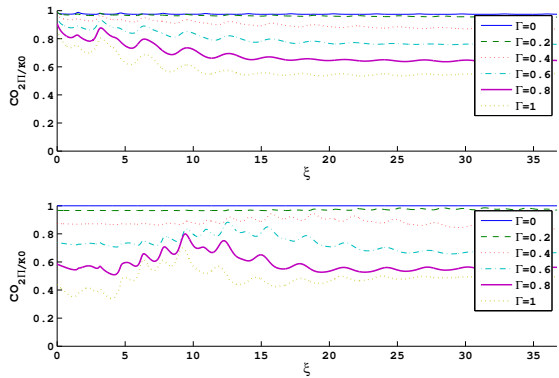}
\noindent\newline
\end{minipage}
\caption{Correlator $\mathrm{CO}_{2\protect\pi /K_{0}}$, characterizing the
proximity of the SP pair to a periodically oscillating mode, for different
strengths of nonlinearity. The top and bottom panels pertain to narrow and
broad pulses, with $T_{0}=2.5$ and $T_{0}=20$, respectively. }
\label{fig5:Fig5}
\end{figure}

To explore the effect of the nonlinearity in a systematic way, we computed
the correlator $\mathrm{CO}_{2\pi /K_{0}}$, using the numerical solutions of
the full system (\ref{eqn:14}). The results, shown in Fig. \ref{fig5:Fig5},
make it evident that stronger nonlinearity worsens the stability of the SP
pair. It is worthy to note too that narrower pulses, with smaller $T_{0}$,
are more robust against the action of the nonlinearity, which is explained
by the similarity of the narrow SPs to the DM solitons, as well as to
stationary gap soliton, which exist in the same system \cite{Kaup}.

\textit{Conclusions}. We have demonstrated that robust periodically
breathing SP (solitary-pulse) pairs can be constructed by linearly coupling
two modes with opposite GVD coefficients. In the linear system, a virtually
exact analytical solution is found by means of the VA for broad pulses, near
edges of the spectral gap. This solution is formally identical to one in the
DM\ model. The VA works well for broad SPs in the nonlinear system as well,
reducing the coupled NLS equations to a single one, which includes both the
DM and nonlinearity management. Strong nonlinearity tends to destabilize the
periodic oscillatory dynamics, although narrower solitons may be
sufficiently robust in the nonlinear system. It may be interesting to extend
the analysis for higher-order modes, such as dipole SPs.

%

\end{document}